\documentclass[twocolumn,prx,superscriptaddress,amsmath,amssymb]{revtex4}

\usepackage{booktabs}
\usepackage{color}
\usepackage{graphicx}% Include figure files
\usepackage{dcolumn}% Align table columns on decimal point
\usepackage{bm}% bold math
\usepackage{graphicx}
\usepackage[colorlinks,linkcolor=blue]{hyperref}
\begin{document}

\title{An Overview of Compton Scattering Calculation}

\author{Chen-Kai Qiao}
%\thanks{This is the corresponding author}
\email{chenkaiqiao@cqut.edu.cn}
\thanks{\\ This is the corresponding author.}
\affiliation{College of Science, Chongqing University of Technology, Banan, Chongqing, 400054, China}

\author{Lin Chen}
\affiliation{College of Science, Chongqing University of Technology, Banan, Chongqing, 400054, China}

\author{Jian-Wei Wei}
\affiliation{College of Science, Chongqing University of Technology, Banan, Chongqing, 400054, China}

\date{\today}

\begin{abstract}
The Compton scattering process plays significant roles in atomic and molecular physics, condensed matter physics, nuclear physics and material science. It could provide useful information on the electromagnetic interaction between light and matter. Several aspects of many-body physics, such us electronic structures, electron momentum distributions, many-body interactions of bound electrons, \emph{etc}, can be revealed by Compton scattering experiments. In this work, we give a review on \emph{ab initio} calculation of Compton scattering process. Several approaches, including the free electron approximation (FEA), impulse approximation (IA), incoherent scattering function / incoherent scattering factor (ISF) and scattering matrix (SM) are focused in this work. The main features and available ranges for these approaches are discussed. Furthermore, we also briefly introduce the databases and applications for Compton scattering.

\

\noindent 
\bf{key words:}

\noindent
Compton scattering, bound electron, many-body interaction, \emph{ab initio} approach
\end{abstract}

\pacs{..}

\maketitle

%\tableofcontents

\section{Introduction \label{sec:1}}

\begin{figure}
	\includegraphics[width=0.5\textwidth]{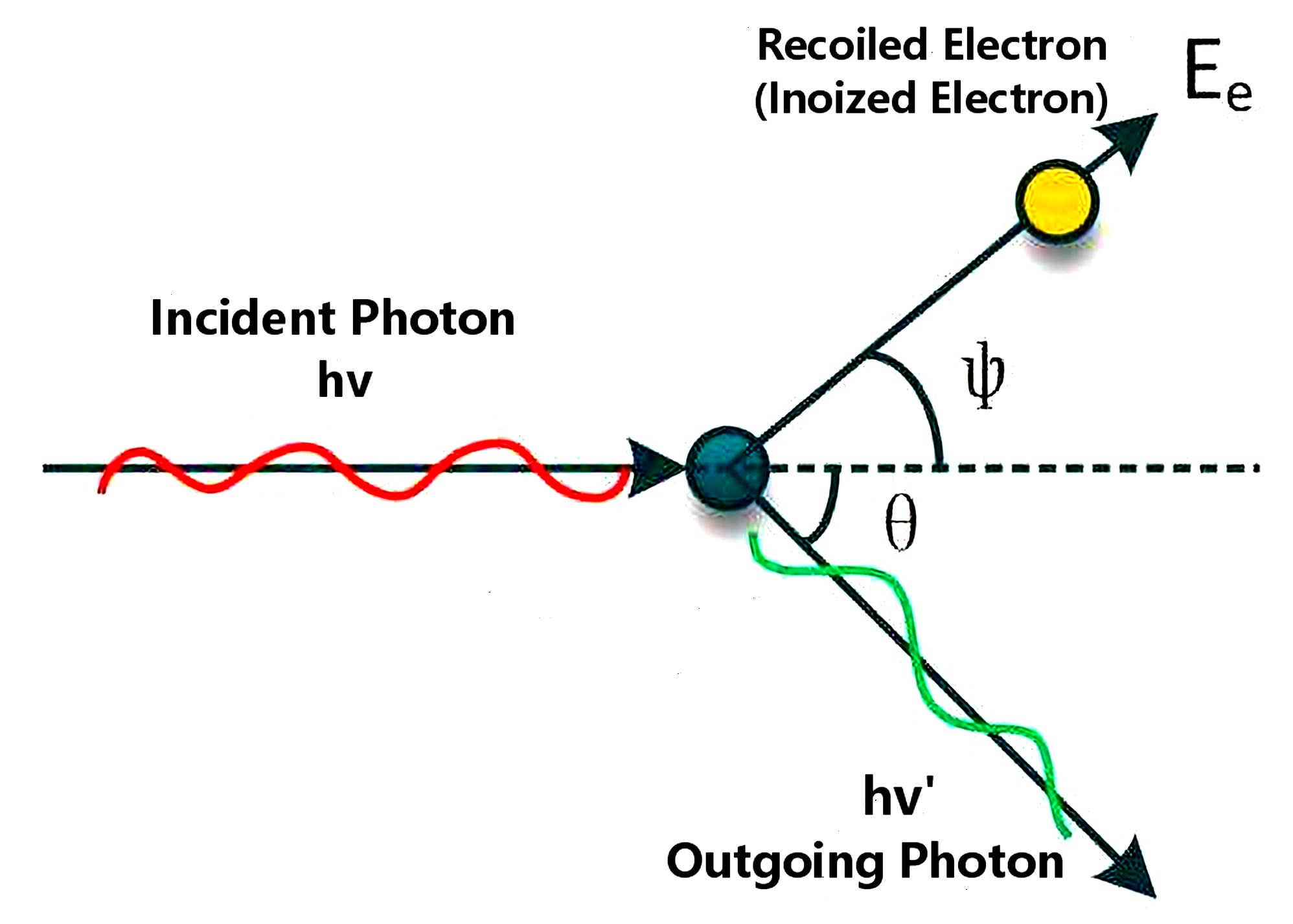}
	\caption{Schematic plot of Compton scattering process $e+\gamma \rightarrow e+\gamma'$.}
	\label{Energy Spectrum}
\end{figure}

Compton scattering process is the scattering between a bound electron ($e$) in atomic or molecular system and an incident photon ($\gamma$) in the electromagnetic field
\[
e+\gamma \rightarrow e+\gamma'
\]
It is one of the most important and mysterious electromagnetic processes in physics. This process was first noticed by A. H. Compton in 1923 \cite{Compton1,Compton2}, through which the quantum nature of X-rays was revealed successfully. Since Compton scattering was discovered in 1920s, it has been carefully studied and extensively investigated for almost a century. The study of atomic Compton scattering could give us information on the interaction between light and matter, and it can also provide opportunities to reveal the underling nature of electric structures, electron correlations, electron momentum distributions, and other aspects of physics \cite{Cooper1,Cooper2,Cooper3}. Because of widely applications, Compton scattering is always an attractive topic in atomic physics, molecular physics, condensed matter physics, nuclear physics and material science \cite{Cooper3,Kubo,Rathor,Wang,Pisani}. (See figure \ref{Energy Spectrum})

Since 1920s, great efforts has been made to develop theoretical methods on \emph{ab initio} calculations for Compton scattering process. The simplest approach is the free electron approximation (FEA), which was first developed by O. Klein and Y. Nishina in 1929 \cite{Klein-Nishina}. In the FEA approach, the electrons in atomic or molecular systems are treated as free electrons. The atomic binding, electron shielding, electron correlations, electron motions around atomic nucleus, and other many-body effects are all omitted for simplicity. In FEA, the angular distribution for Compton scattering process is given by the Klein-Nishina formula \cite{Klein-Nishina,Sakurai}
\begin{equation}
	\bigg(\frac{d\sigma}{d\Omega_{f}}\bigg)_{\text{FEA}}
	=\frac{r_{0}^{2}}{2}
	\bigg(
	\frac{\omega_{C}}{\omega_{i}}
	\bigg)^{2}
	\bigg(
	\frac{\omega_{i}}{\omega_{C}}+\frac{\omega_{C}}{\omega_{i}}-\sin^{2}\theta
	\bigg)
	\label{FEA}
\end{equation}
Furthermore, in FEA approach, the final state photon energy $\omega_{f}$ after the scattering process is totally determined by the scattering angle $\theta$ in Compton scattering process
\begin{equation}
	\omega_{f}=\omega_{C}=\frac{\omega_{i}}{1+\omega_{i}(1-\cos\theta)/m_{e}c^{2}}
\end{equation}
Here, $\omega_{C}$ is called as the Compton energy. Because of computational simplicity and clearly physical meaning, the FEA approach is most widely used in many branches of science. over the past few decades, it has become a standard treatment for atomic Compton scattering in many textbooks. However, the FEA approach is much too simple, very little information on electronic structures and property of materials can be acquired. From equation (\ref{FEA}), it is evident that the angular distribution $\big(d\sigma/d\Omega_{f}\big)_{\text{FEA}}$ is independent of the electron momentum distributions in target materials.

Soon after the FEA approach was formulated by O. Klein and Y. Nishina in 1929, other approaches considering bound structures in atoms and molecules emerged. It was J. W. M. DuMond first realized that the atomic effects and electron motions would give inevitable effects on the scattering process. Then he introduced an approach to effectively include atomic binding effects as well as electron motion around atomic nuclei in the calculation of Compton scattering \cite{DuMond1,DuMond2,DuMond3}. His approach was later called as the impulse approximation (IA). The DuMond's approach is the nonrelativistic impulse approximation (NRIA), and the same result was re-derived by P. Eisenberger \emph{et al.} in 1970s \cite{Eisenberger1,Eisenberger2,Eisenberger3,Currat}. The relativistic impulse approximation (RIA) was developed until 1970s-1980s by R. Ribberfors \emph{et al.} \cite{Eisenberger3,Ribberfors1,Ribberfors2,Ribberfors3,Ribberfors4,Ribberfors5}. In the IA formulation, the doubly differential cross section (DDCS) for Compton scattering process can be factorized into two parts as follows:
\begin{equation}
	\bigg( \frac{d^{2}\sigma}{d\omega_{f}d\Omega_{f}} \bigg)_{\text{IA}} = Y_{\text{IA}} \cdot J(p_{z}) \label{IA}
\end{equation}
where $\Omega_{f}$ is the solid angle for final state outgoing photon. In equation (\ref{IA}), $Y_{\text{IA}}$ is a factor dependent on kinematical and dynamical properties of atomic Compton scattering, and irrelevant to the electronic structure of target materials. The factor $J(p_{z})$, which is known as the Compton profile, is related to the momentum distributions of bound electrons in the atomic or molecular systems \cite{Biggs}. The Compton profile can reflect electronic structures, physical and chemical properties of target materials. It can be determined from theoretical calculations and experimental measurements. In actual \emph{ab initio} calculations, the Compton profile $J(p_{z})$ can be obtained using the nonrelativistic Hartree-Fock theory (HF), the relativistic Dirac-Hartree Fock theory (DHF), and the density functional theory (DFT). It is worth noting that, in the IA approach, all the many-body effects in atomic and molecular systems in Compton scattering processes are incorporated into Compton profiles $J(p_{z})$.

From the IA approach, the angular distribution of Compton scattering process can be obtained by integration of equation (\ref{IA}), and the results is also a simple correction for FEA:
\begin{equation}
	\bigg( \frac{d\sigma}{d\Omega_{f}} \bigg)_{\text{IA / ISF}} = \bigg( \frac{d\sigma}{d\Omega_{f}} \bigg)_{\text{FEA}} SF(\omega_{i},\theta) \label{RIA2}
\end{equation}
Here, $\big( d\sigma/d\Omega_{f} \big)_{\text{FEA}}$ is the angular distribution calculated in FEA approach as shown in equation (\ref{FEA}), and the correction factor $SF(\omega_{i},\theta)$ is called the scattering function. It should be mentioned that, apart from IA, there are other approaches which can give the same results in equation ({\ref{RIA2}}). In other words, scattering function $SF(\omega_{i},\theta)$ can also be calculated by other methods, such as the nonrelativistic Waller-Hartree theory \cite{Waller,Freeman1959}. This kind of approach, in which the angular distribution of Compton scattering is given by equation (\ref{RIA2}), is called the incoherent scattering function or incoherent scattering factor (ISF) approach. Same as Compton profile $J(p_{z})$, the scattering function $SF(\omega_{i},\theta)$ can also reflect the electronic properties of target materials. On the one hand, $J(p_{z})$ and $SF(\omega_{i},\theta)$ can be measured from Compton scattering experiments with high precision. On the other hand, they can also be predicted by theoretical \emph{ab initio} calculations in atomic, molecular, and condensed matter physics \cite{Biggs,Hubbell}. The Compton profile $J(p_{z})$ and scattering function $SF(\omega_{i},\theta)$ provide a bridge between Compton scattering and interdisciplinary studies in many branches of science. They can offer opportunities to learn the electronic structure and properties of target materials from Compton scattering process. Recently, there are many studies in which electronic structures and properties (include electron interactions, electron correlations, electron momentum distributions, band structures and fermi surfaces) are investigated through Compton profile and scattering function \cite{Kubo,Pisani,Rathor,Wang}.

In the past decades, several methods beyond IA and ISF were emerged and formulated \cite{Pratt1,Pratt,Kaplan,Suric,Pratt2,Drukarev1,Drukarev2}. Among these approaches, the most successful one is based on the perturbation expansion of many-body electromagnetic interactions in atomic or molecular systems. In this approach, the DDCS of Compton scattering process can be calculated by the scattering matrix of a many-body theory
\begin{equation}
	\bigg( \frac{d^{2}\sigma}{d\omega_{f}d\Omega_{f}} \bigg)_{\text{SM}} \propto \mid M_{if} \mid^{2} \label{SM}
\end{equation}
This approach is called the Scattering Matrix (SM) approach. In the SM, scattering matrix of Compton scattering process $M_{if} \propto \langle\Psi_{f}|H_{I}|\Psi_{i}\rangle$ can be calculated through the many-body interaction Hamiltonian $H_{I}$, which gives the many-body interaction between photon and bound electrons in atomic or molecular systems. The SM approach can take into account the factor of atomic bindings and electron interactions in atomic or molecular systems as much as possible. SM could handle with these many-body effects precisely in the dynamical process of Compton scattering. Recently, the SM formulation has revealed many nontrivial properties of Compton scattering, and it has attracted lots of interests in interdisciplinary studies \cite{Pratt,Pratt2}.

Furthermore, there are several experiments which give supporting to the SM predictions \cite{Sparks,Kane1997,Jung1998,Pratt2000,Pratt2004,Chatterjee,Kircher}. Recently, Max Kircher \emph{et al.} conducted a kinematically complete Compton scattering experiment utilizing X-rays produced from accelerators with energy about 2.1 keV. By measuring the angular distribution of scattered photons, the experimental observations present large deviations with the FEA results. However, when SM approach is employed and \emph{ab initio} numerical calculations are carried out, the experimental data are consistent with theoretical predictions \cite{Kircher}. These results indicate that the SM approach is becoming an promising tool to duel with Compton scattering with bound electrons, and it may have great impacts in this area in the near future.

This paper is organized as follows. Section \ref{sec:1} gives an introduction on Compton scattering and the development of theoretical methods on \emph{ab initio} calculations of Compton scattering process. Section \ref{sec:2} gives a description of FEA, section \ref{sec:3} discuss the IA approach, and section \ref{sec:4} is devoted to the ISF. The most advanced SM approach is presented in section \ref{sec:5}. Section \ref{sec:6} presents the comparisons between theoretical calculations and experimental measurements. Databases and applications for Compton scattering are briefly introduced in section \ref{sec:7}. Summaries are presented in section \ref{sec:8}.

\section{Free Electron Approximation \label{sec:2}}

In the free electron approximation (FEA), the electrons in Compton scattering are treated as free electrons, all the atomic binding effects and many-body interactions of bound electrons in atomic or molecular systems are neglected in the scattering process. Further, it is also assumed that the bound electrons are at rest before the Compton scattering.

The FEA approach for Compton scattering process was given by O. Klein and Y. Nishina in 1929 \cite{Klein-Nishina}. In this formulation, the scattered photon energy $\omega_{f}$ in Compton scattering process is totally determined by its scattering angle $\theta$ via equation
\begin{equation}
	\omega_{f}=\omega_{C}=\frac{\omega_{i}}{1+\omega_{i}(1-\cos\theta)/m_{e}c^{2}} \label{Compton energy}
\end{equation}
Here, $m_{e}$ is the mass of electron, and $\omega_{C}$ is called as the Compton energy. When $\theta=180^{\text{o}}$, the energy of scattered photon $\omega_{f}$ reaches its minimum, meanwhile the energy transfer $T=\omega_{i}-\omega_{C}$ arrives at its maximum. They can be expressed in the following
\begin{eqnarray}
	\omega_{C}^{\text{min}} & = & \omega_{f}^{\text{min}}=\frac{\omega_{i}}{1+2\omega_{i}/m_{e}c^{2}}
	\\
	T^{\text{max}} & = & \omega_{i}-\omega_{C}^{\text{min}}=\omega_{i}-\frac{\omega_{i}}{1+2\omega_{i}/m_{e}c^{2}}
\end{eqnarray}
They correspond to the Compton edge of the Compton scattering spectrum of $d\sigma/dT$ \cite{Qiao,Qiao2}.

In FEA formulation, the angular distribution of Compton scattering process is given by the Klein-Nishina formula \cite{Klein-Nishina,Sakurai}
\begin{equation}
	\bigg(
	\frac{d\sigma}{d\Omega_{f}}
	\bigg)_{\text{FEA}}
	=
	\frac{r_{0}^{2}}{2}
	\bigg(
	\frac{\omega_{C}}{\omega_{i}}
	\bigg)^{2}
	\bigg(
	\frac{\omega_{i}}{\omega_{C}}+\frac{\omega_{C}}{\omega_{i}}-\sin^{2}\theta
	\bigg) \label{KN}
\end{equation}
and the corresponding DDCS can be expressed as
\begin{eqnarray}
	\bigg(
	\frac{d^{2}\sigma}{d\omega_{f}d\Omega_{f}}
	\bigg)_{\text{FEA}}
	& = &
	\frac{r_{0}^{2}}{4}
	\bigg(
	\frac{\omega_{f}}{\omega_{i}}
	\bigg)^{2}
	\bigg[
	\frac{\omega_{i}}{\omega_{f}}+\frac{\omega_{f}}{\omega_{i}}
	+4(\boldsymbol{\epsilon}_{i}\cdot\boldsymbol{\epsilon}_{f})^2-2
	\bigg] \nonumber
	\\
	&   &
	\times \delta(\omega_{f}-\omega_{C})
	\label{KN doubly-differential}
\end{eqnarray}
Here, $r_{0}$ is the classical radius of electron, $\boldsymbol{\epsilon}_{i}$ and $\boldsymbol{\epsilon}_{f}$ are polarization vectors for incoming and outgoing photons, $\omega_{f}$ and $\Omega_{f}$ are the energy and solid angle of the outgoing scattered photon, respectively. In the FEA results, due to the Dirac delta function $\delta(\omega_{f}-\omega_{C})$ in equation (\ref{KN doubly-differential}), the spectrum of DDCS is an isolated line located at Compton energy $\omega_{C}$, which is named as ``Compton line'', as illustrated in figure \ref{Compton Scattering Distribution}.

In the nonrelativistic limit, where the incoming photon energy $\omega_{i} \ll m_{e}c^{2}$, the angular distribution of Compton scattering reduced to
\begin{equation}
	\bigg(
	\frac{d\sigma}{d\Omega_{f}}
	\bigg)_{\text{FEA}}
	=
	\frac{r_{0}^{2}}{2}
	\bigg(
	\frac{\omega_{C}}{\omega_{i}}
	\bigg)^{2}
	\bigg(
	1+\cos^{2}\theta
	\bigg)
	\label{NRFEA}
\end{equation}
In this case, the Compton energy becomes
\begin{eqnarray}
	\omega_{f}=\omega_{C} & = & \frac{\omega_{i}}{1+\omega_{i}(1-\cos\theta)/m_{e}c^{2}} \nonumber
	\\
	& \approx & \omega_{i}
	\bigg[ 1-\frac{\omega_{i}}{m_{e}c^{2}}(1-\cos\theta) \bigg] \nonumber
	\\ & \approx & \omega_{i}-\frac{K^{2}}{2m_{e}}
\end{eqnarray}
where $K$ is the momentum transfer in the Compton scattering process. Furthermore, in the elastic scattering limit $\omega_{f}=\omega_{C} \rightarrow \omega_{i}$ (namely the scattering angle $\theta \rightarrow 0$), the equation (\ref{NRFEA}) further reduce to the Thomson formula
\begin{equation}
	\bigg(
	\frac{d\sigma}{d\Omega_{f}}
	\bigg)_{\text{Thomson}}
	=
	\frac{r_{0}^{2}}{2}
	\bigg(
	1+\cos^{2}\theta
	\bigg)
\end{equation}
This is the differential cross section of elastic scattering process between photon and free electron \cite{Sakurai,Thomson,Biggs,Hubbell}.
It should be mentioned that the Klein-Nishina formula in equation (\ref{KN}) in FEA works well only in the cases that atomic binding energies are negligible and the atomic electrons are approximately free. When the incident photon energy $\omega_{i}$ and energy transfer $T$ are extremely high, the FEA result is appropriate. However, in many cases, the incident photon energy is comparable to the X-ray characteristic energies, in which the atomic bindings and electron many-body interactions cannot be neglected. Then the FEA formulation becomes invalid and it fails to fit the experimental data \cite{Pratt1}.

\section{Impulse Approximation \label{sec:3}}

In the impulse approximation (IA) method, the atomic binding effects are effectively considered, and the electron pre-collision motions around atomic nuclei are also included. The basic starting point of IA approach can be shown in the following way. Because of atomic binding and electrons many-body interactions, the bound electrons in atomic and molecular systems have a momentum distribution $\rho(\boldsymbol{p})$ when moving around atomic or molecular nuclei. In principle, the momentum distribution for bound electrons is determined by ground state wavefunctions in atomic or molecular systems. In the Compton scattering process, suppose electron in momentum eigenstate $|\boldsymbol{p}\rangle$ scattered with incoming photon $\gamma$ very rapidly, like an impulse acting on the electron. This scattering process is too quick to be disturbed by other electrons. In this way, electron momentum eigenstate $|\boldsymbol{p}\rangle$ scattered with incident photon $\gamma$ independently as free electrons. Many-body interactions and interference terms between electrons with different momentum eigenstates ($|\boldsymbol{p}\rangle$ and $|\boldsymbol{p'}\rangle$) in the dynamical process of Compton scattering are omitted for simplicity. Therefore, the DDCS of the Compton scattering process is achieved through a summation of the scattering probability for all possible momentum eigenstates
\begin{eqnarray}
	\bigg( \frac{d^{2}\sigma}{d\omega_{f}d\Omega_{f}} \bigg)_{\text{IA}}
	& = &
	\iiint{d^{3}p\rho(\boldsymbol{p})
		\bigg( \frac{d^{2}\sigma(\boldsymbol{p})}{d\omega_{f}d\Omega_{f}} \bigg)_{\text{FEA}}} \nonumber
	\\
	&   &
	\times \ \delta (E_{i}+\omega_{i}-E_{f}-\omega_{f}) \label{IA Starting}
\end{eqnarray}
In this expression, $\big( d^{2}\sigma(\boldsymbol{p})/d\omega_{f}d\Omega_{f} \big)_{\text{FEA}}$ is the DDCS of Compton scattering between photon $\gamma$ and electron in momentum eigenstate $|\boldsymbol{p}\rangle$, which can be calculated by FEA. The $E_{i}$ and $E_{f}$ are the energies of electron before and after Compton scattering process. The equation (\ref{IA Starting}) can be viewed as the basic starting point of IA approach, and different versions of IA treatments are achieved by applying different numerical schemes in calculating equation (\ref{IA Starting}) \cite{Qiao}.

\begin{figure}
	\includegraphics[width=0.525\textwidth]{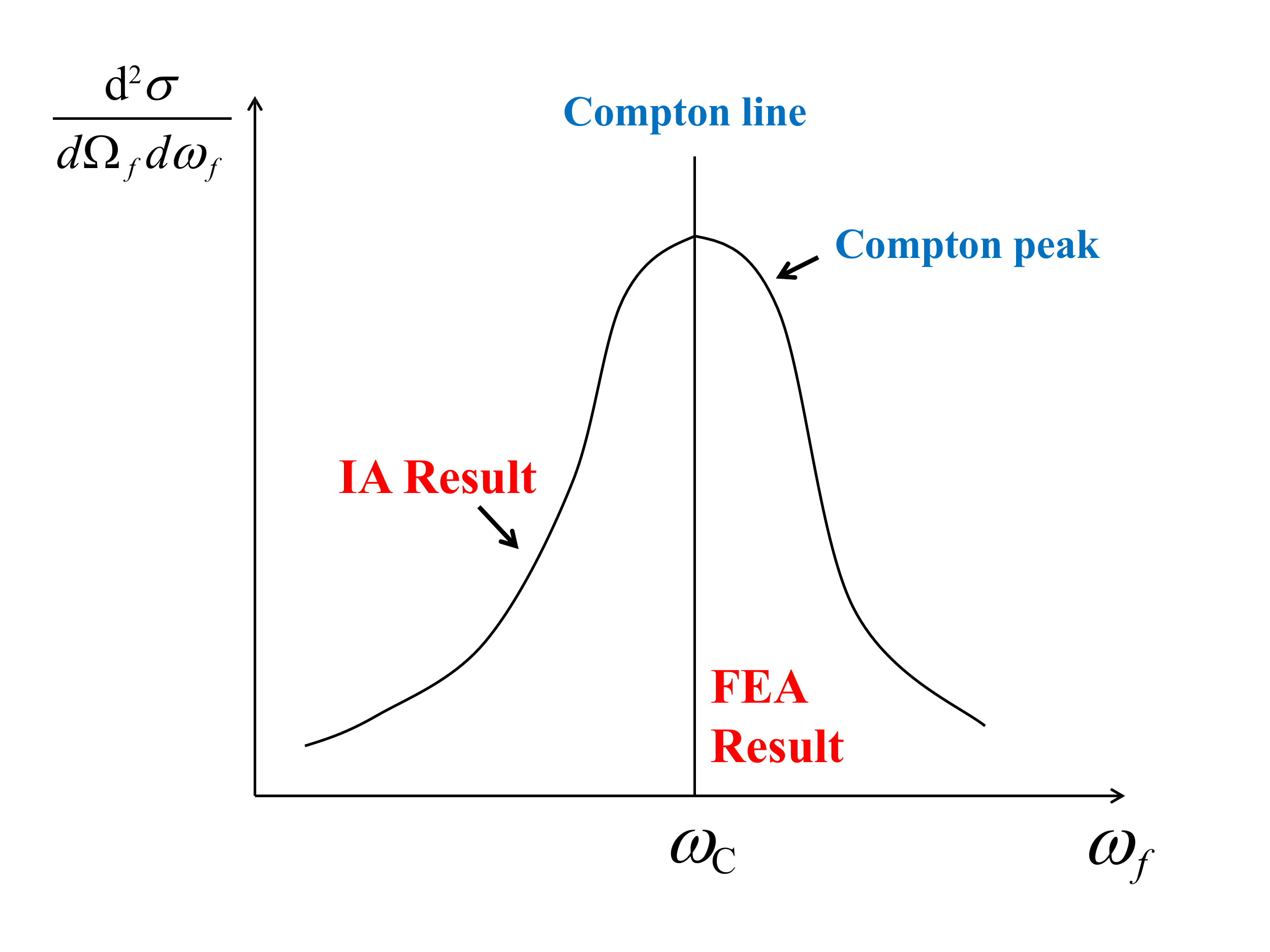}
	\caption{Schematic plot of the FEA and IA results of DDCS at scattering angle $\theta$. The horizontal axis labels the finial photon energy $\omega_{f}$, and the vertical axis labels the DDCS of Compton scattering process.}
	\label{Compton Scattering Distribution}
\end{figure}

The aforementioned atomic binding effects and electron pre-collision motion make the IA results significantly different from FEA results. In the FEA results, there is a one-to-one correspondence between the final state photon energy $\omega_{f}$ and the scattering angle $\theta$ (see equation (\ref{Compton energy})). While in the IA approach, due to the electron motion around atomic nuclei, the energy of the scattered photon $\omega_{f}$ can not be totally determined by its scattering angle $\theta$ as in FEA. For the same scattering angle $\theta$, the outgoing photon energy $\omega_{f}$ has a continuous distribution. The maximum cross section is located at the Compton energy $\omega_{C}$, which is given by equation (\ref{Compton energy}), forming a ``Compton peak'' in the spectrum of DDCS. In figure \ref{Compton Scattering Distribution}, the DDCS of Compton scattering predicted in FEA and IA approaches are illustrated schematically. Sometimes, it is more enlightening to interpret the emergence of ``Compton peak'' as the Doppler broadening effect of the ``Compton line'' (located at Compton energy $\omega_{C}$) \cite{Namito1994,Ordonez1997}. This Doppler effect is due to the bound electrons' motion around atomic nuclei.

The nonrelativistic impulse approximation (NRIA) was first developed by J. W. M. DuMond in the 1930s \cite{DuMond1,DuMond2,DuMond3}. In 1970s, P. Eisenberger and P. M. Platzman re-derived the NRIA formulation based on the more accurate scattering matrix (SM) approach. It is shown that NRIA can be viewed as the leading order approximation of SM approach \cite{Eisenberger1,Eisenberger2,Eisenberger3,Currat}. In NRIA, the DDCS of Compton scattering is given by \cite{DuMond3,Eisenberger1}
\begin{eqnarray}
	\bigg(
	\frac{d^{2}\sigma}{d\omega_{f}d\Omega_{f}}
	\bigg)_{\text{NRIA}}
	& = &
	\frac{r_{0}^{2}}{2}\frac{m_{e}}{K}
	\frac{\omega_{f}}{\omega_{i}}
	(1+\cos^{2}\theta)
	J(p_{z}) \nonumber
	\\
	& = &
	Y_{\text{NRIA}} \cdot J(p_{z})
	\label{NRIA}
\end{eqnarray}
Here, $K$ is the modulus of the momentum transfer vector $\boldsymbol{K} \equiv \boldsymbol{k}_{f}-\boldsymbol{k}_{i}$ in Compton scattering process, and $p_{z}$ is given by
\begin{equation}
	p_{z}=\frac{K}{2}-\frac{m_{e}(\omega_{i}-\omega_{f})}{K}
\end{equation}
In equation (\ref{NRIA}), the factor $Y_{\text{NRIA}}=r_{0}^{2}m_{e}\omega_{f}(1+\cos^{2}\theta)/2K\omega_{i}$ relies on the dynamical and kinematical properties of Compton scattering process in the nonrelativistic limit. The $J(p_{z})$ is a factor coming from the many-body effects in atomic or molecular systems, which is named as the ``Compton profile'' \cite{Biggs}
\begin{equation}
	J(p_{z})\equiv \iint\rho(\boldsymbol{p})dp_{x}dp_{y} \label{Compton profile}
\end{equation}
where $\rho(\boldsymbol{p})$ is the electron momentum density of the atomic or molecular ground-sataes. The Compton profile is closely related to the electronic properties of the atomic or molecular systems, and it has been widely studied in atomic, molecular and condensed matter physics \cite{Kubo,Rathor,Pisani}. Furthermore, for most of the atomic systems, the momentum distribution is spherical symmetric, then the Compton profile reduces to
\begin{equation}
	J(p_{z})=2\pi\int\limits_{|p_{z}|}^{\infty}p\rho(p)dp
	\label{electron profile2}
\end{equation}
In these cases, the Compton profile $J(p_{z})$ is bell-shaped and axisymmetric around the $p_{z}=0$ axis. However, many of the molecular or condensed matter systems do not have this spherical symmetric property.

The relativistic impulse approximation (RIA) was developed by R. Ribberfors \emph{et al.} in 1975-1985 \cite{Eisenberger3,Ribberfors1,Ribberfors2,Ribberfors3,Ribberfors4,Ribberfors5}. In this formulation, the DDCS of Compton scattering is given by \cite{Ribberfors3,Ribberfors4,Brusa}
\begin{eqnarray}
	\bigg(
	\frac{d^{2}\sigma}{d\omega_{f}d\Omega_{f}}
	\bigg)_{\text{RIA}}
	& = &
	\frac{r_{0}^{2}}{2}\frac{m_{e}}{K}
	\frac{m_{e}c^{2}}{E(p_{z})}
	\frac{\omega_{f}}{\omega_{i}}
	\overline{X}(p_{z})
	J(p_{z}) \nonumber
	\\
	& = &
	Y_{\text{NRIA}} \cdot J(p_{z})
	\label{RIA}
\end{eqnarray}
The same as in NRIA, $Y_{\text{RIA}}=r_{0}^{2}m_{e}^{2}c^{2}\omega_{f}\overline{X}(p_{z})/2K\omega_{i}E(p_{z})$ is a factor depends on kinematical and dynamical properties of Compton scattering. The factor relevant to atomic or molecular structure is the Compton profile $J(p_{z})$ defined in equation (\ref{Compton profile}). In equation (\ref{RIA}), $K$ is the modulus of the momentum transfer vector $\boldsymbol{K} \equiv \boldsymbol{k}_{f}-\boldsymbol{k}_{i}$ and $p_{z}$ is the projection of the electron's initial momentum on the momentum transfer direction
\begin{equation}
	p_{z} = \frac{\boldsymbol{p}\cdot\boldsymbol{K}}{K}
	= \frac{\omega_{i}\omega_{f}(1-\cos\theta)-E(p_{z})(\omega_{i}-\omega_{f})}{c^{2}K} \label{projection momentum}
\end{equation}
Moreover, the function $\overline{X}(p_{z})$ is defined to be
\begin{eqnarray}
	\overline{X}(p_{z})
	& = &
	\frac{K_{i}(p_{z})}{K_{f}(p_{z})}
	+\frac{K_{f}(p_{z})}{K_{i}(p_{z})} \nonumber
	\\
	&   &
	+2m_{e}^{2}c^{2}
	\bigg(
	\frac{1}{K_{i}(p_{z})}-\frac{1}{K_{f}(p_{z})}
	\bigg) \nonumber
	\\
	&   &
	+m_{e}^{4}c^{4}
	\bigg(
	\frac{1}{K_{i}(p_{z})}-\frac{1}{K_{f}(p_{z})}
	\bigg)^{2} 
	\label{function X Pmin}
\end{eqnarray}
with $K_{i}$ and $K_{f}$ defined as
\begin{eqnarray}
	K_{i}(p_{z}) & = & \frac{\omega_{i}E(p_{z})}{c^{2}}+\frac{\omega_{i}(\omega_{i}-\omega_{f}\cos\theta)p_{z}}{c^{2}q}
	\\
	K_{f}(p_{z}) & = & K_{i}(p_{z})-\frac{\omega_{i}\omega_{f}(1-\cos\theta)}{c^{2}}
	\\
	E(p_{z}) & = & \sqrt{m_{e}^{2}c^{4}+p_{z}^{2}c^{2}}
\end{eqnarray}

To summarize, combining equations (\ref{NRIA}) and (\ref{RIA}), it is clearly that the DDCS of Compton scattering process calculated in IA approach factorizes into two parts
\begin{equation}
	\bigg( \frac{d^{2}\sigma}{d\omega_{f}d\Omega_{f}} \bigg)_{\text{IA}} = Y_{\text{IA}} \cdot J(p_{z})
\end{equation}
where $Y_{\text{IA}}$ is a factor dependent on kinematical and dynamical properties of Compton scattering process, and the Compton profile $J(p_{z})$ is relevant to the electron momentum distribution of the atomic or molecular ground-states. It is worth noting that, in the IA approach, all the many-body effects in atomic and molecular systems in Compton scattering process are incorporated into Compton profiles $J(p_{z})$. The Compton profile can reflect electronic structures and properties of target materials. Therefore, it provide us an opportunity to learn the electronic structures, electron interactions and properties of materials from Compton scattering processes. In actual \emph{ab initio} calculations, the Compton profile $J(p_{z})$ can be calculated by the nonrelativistic Hartree-Fock theory (HF), the relativistic Dirac-Hartree Fock theory (DHF), and the density functional theory (DFT) \cite{Biggs,Biggs1974,Grant,Bauer1983a,Bauer1983b,Parr1986,Aguiar2015}. Interestingly, the FEA result can be simply reduced from IA result by replacing the Compton profile with the Dirac delta distribution $J(p_{z})=\delta(p_{z})$. Considering the atomic binding effects and electron pre-collision motions, the IA formulation could overcome the shortcomings in the FEA method. It is a practical approach to calculate the Compton scattering of bound electrons with X-rays and gamma-rays. Later researches shown that the IA approach could serve as a good approximation in energy region near the ``Compton peak'' \cite{Pratt,Qiao} (see figure \ref{Compton Scattering Distribution} and figure \ref{Compton Scattering Distribution SM}).

Although the IA formulation effectively takes into consideration the atomic binding effects and electron pre-collision motions around atomic nuclei, it still has limitations in dealing with Compton scattering. In IA formulation, all the many-body effects in Compton scattering process are incorporated into Compton profile $J(p_{z})$, which is an observable only related to the atomic or molecular ground-states. The many-body effects coming from the ionized states and the interference terms among different momentum eigenstates are still insufficient in the dynamical process of Compton scattering. In the past few years, several approaches beyond the IA had already been investigated \cite{Pratt1,Pratt,Kaplan,Suric,Pratt2,Drukarev1,Drukarev2}. These researches, which mainly employ the low-energy theorem (LET) and scattering matrix (SM) approaches, have revealed many nontrivial properties of Compton scattering and have attracted lots of interests in interdisciplinary studies. Through comparing IA with these more advanced approaches, it is clearly that the validity region for IA approach is just near the Compton peak region.  Further, in the validity region of IA, the momentum transfer $K$ in Compton scattering is much larger than the average momentum $p_{\text{average}}$ for bound electrons (namely $p_{\text{average}}/K \ll 1$) \cite{Pratt1,Pratt2,Bergstrom,Pratt2011}. We will specialize in the SM approach in section \ref{sec:5}.

\section{Incoherent Scattering Factor / Incoherent Scattering Function \label{sec:4}}

In this section, we give an introduction of the incoherent scattering function / incoherent scattering factor (ISF) approach. First, we can use the IA result as an example to illustrate the ISF approach. Starting from the IA result of DDCS in equations (\ref{NRIA}) and (\ref{RIA}), the angular distribution of Compton scattering can be calculated by an integration of final state photon energy $\omega_{f}$ over the allowed range
\begin{equation}
	\frac{d\sigma}{d\Omega_{f}}=\int_{\omega_{\text{min}}}^{\omega_{\text{max}}}\frac{d^{2}\sigma}{d\omega_{f}d\Omega_{f}}d\omega_{f} \label{RIA differential}
\end{equation}
After this integration, the angular distribution $d\sigma/d\Omega_{f}$ can be reduced to the product of Klein-Nishina result in equation (\ref{KN}) and a correction factor \cite{Ribberfors3,Ribberfors4}.
\begin{equation}
	\bigg( \frac{d\sigma}{d\Omega_{f}} \bigg)_{\text{IA / ISF}} = \bigg( \frac{d\sigma}{d\Omega_{f}}\bigg)_{\text{FEA}} SF(\omega_{i},\theta) \label{SF differential}
\end{equation}
The correction factor $SF(\omega_{i},\theta)$ is called the scattering function or scattering factor. Apart from IA, there are other ways to give the same results in equation (\ref{SF differential}) \cite{Waller,Freeman1959}. This kind of approach, in which the angular distribution of Compton scattering is given by equation (\ref{SF differential}) and $SF(\omega_{i},\theta)$, is called the ISF approach.

We can use the following way to understand the physical meaning of scattering function $SF(\omega_{i},\theta)$. If the atomic or molecular system has N electrons, and each electron scattered with photon independently as free electrons. Then the angular distribution of Compton scattering for the whose system becomes
\begin{equation}
	\bigg( \frac{d\sigma}{d\Omega_{f}} \bigg)^{\text{total system}}_{\text{FEA}}
	= N \bigg( \frac{d\sigma}{d\Omega_{f}}\bigg)_{\text{FEA}}^{\text{single electron}}
\end{equation}
Therefore, the scatting function $SF(\omega_{i},\theta)$, which is defined to be the ratio between the contribution of total system and that of single electron
\begin{equation}
	SF(\omega_{i},\theta) \equiv
	\frac{\big( d\sigma/d\Omega_{f} \big)^{\text{total system}} }
	{\big( d\sigma/d\Omega_{f} \big)_{\text{FEA}}^{\text{single electron}}  }
\end{equation}
can be view as the number of activated electrons in the Compton scattering process.

Here, we would present the scatting function $SF(\omega_{i},\theta)$ calculated within the RIA approach, the scatting function calculated using NRIA can be obtained in a similar way. To obtain scatting function $SF(\omega_{i},\theta)$ from RIA, we need to substitute the DDCS in equation (\ref{RIA}) into equation (\ref{RIA differential}) and evaluate the integral. After utilizing some approximations, the factorization results of angular distribution in equation (\ref{SF differential}) can be obtained. Finally, the scattering function $SF(\omega_{i},\theta)$ can be expressed as \cite{Ribberfors3,Brusa,Kahane}:
\begin{equation}
	SF(\omega_{i},\theta)=\sum_{i}Z_{i}\Theta(\omega_{i}-U_{i})n_{i}(p_{i}^{\text{max}}) \label{SF RIA}
\end{equation}
The Heaviside step function $\Theta(\omega_{i}-U_{i})$ guarantees that only the activated electrons are included. In Compton scattering process, the electron becomes activated when the transferred energy $T=\omega_{i}-\omega_{f}$ is larger than the binding energy $U_{i}$ for $i$-th subshell. In equation (\ref{SF RIA}), $Z_{i}$ is the number of electron in $i$-th subshell, $p_{i}^{\text{max}}$ denotes the maximum value of $p_{z}$ for the $i$-th subshell electron
\begin{equation}
	p_{i}^{\text{max}}
	=
	\frac{\omega_{i}(\omega_{i}-U_{i})(1-\cos\theta)-m_{e}c^{2}U_{i}}{c\sqrt{2\omega_{i}(\omega_{i}-U_{i})(1-\cos\theta)+U_{i}^2}}
\end{equation}
and function $n_{i}(p_{i}^{\text{max}})$ is defined to be an integral for Compton profile
\begin{equation}\label{ni}
	n_{i}(p_{i}^{\text{max}})=\int\limits_{-\infty}^{p_{i}^{\text{max}}}J_{i}(p_{z})dp_{z}
\end{equation}
Here, the function $J_{i}(p_{z})$ is the single electron Compton profile for $i$-th subshell. It can be expressed as
\begin{equation}
	J_{i}(p_{z})=\iint\rho_{i}(\textbf{p})dp_{x}dp_{y} \label{electron profile}
\end{equation}
The $\rho_{i}(p_{z})$ is the electron momentum distribution for $i$-th subshell, which can be calculated by ground state momentum wavefunctions. Then the total Compton profile for atomic or molecular system is give by
\begin{equation}\label{atomic profile}
	J(p_{z})=\sum_{i}Z_{i}J_{i}(p_{z})
\end{equation}

The quantity $SF(\omega_{i},\theta)$ in equation (\ref{SF RIA}) is the scattering function in RIA formulation. In principle, it is a two-variable function depending on initial photon energy $\omega_{i}$ and scattering angle $\theta$. However, in the nonrelativistic limit, these two variables are related to each other and they can not be fully separated, which makes the scattering function $SF(\omega_{i},\theta)$ further reduce to a single-variable function \cite{Hubbell,Kahane}. The interdependence of $\omega_{i}$ and $\theta$ is realized via a new variable $x$ used in Hubbell's work \cite{Hubbell}
\begin{eqnarray}
	x[{\AA}^{-1}] &=& \sin(\frac{\theta}{2})/\lambda[{\AA}] \label{variable X}
	\\
	\lambda[{\AA}] &=& \frac{12.3984}{\omega_{i}[keV]}
\end{eqnarray}
Actually, in this expression, $x$ is proportional to the momentum transfer in the elastic scattering between photon and electron, which is the Rayleigh scattering process, at scattering angle $\theta$.

It should be mentioned that, although the above ISF result on angular distribution displayed in equation (\ref{SF differential}) is derived from IA approach, the same result can be derived from other theories. In other words, there are alternative methods to calculate the scattering function $SF(\omega_{i},\theta)$ in Compton scattering process. For instance, in the nonrelativistic Waller-Hatree theory, the scattering function is given by \cite{Waller}
\begin{equation}
	SF(K,N)_{\text{WH}}=N-\sum^{N}_{i=1}f_{ii}(K)-\sum^{N}_{i=1}\sum^{N}_{j=1}|f_{ij}(K)|^{2} \label{Waller Hartree_SF0}
\end{equation}
where $N$ is the number of electrons in atomic or molecular systems, and $f_{ij}(K,N)$ is form factor
\begin{equation}
	%	f_{ij}(K)=\iiint\psi_{i}^{*}(\boldsymbol{r})exp(i\mathbf{K}\cdot\mathbf{r})\psi_{j}(\boldsymbol{r})d\mathbf{r} \label{Form Factor}
	f_{ij}(K)=\iiint\psi_{i}^{*}(\boldsymbol{r})e^{i\mathbf{K}\cdot\mathbf{r}}\psi_{j}(\boldsymbol{r})d\mathbf{r} \label{Form Factor}
\end{equation}
Here, $\psi(\boldsymbol{r})$ is the single-electron wavefunction for $i$-th electron, and $\boldsymbol{K} \equiv \boldsymbol{k}_{f}-\boldsymbol{k}_{i}$ is momentum transfer vector. Later, in a work presented by J. H. Hubbell \emph{et al.}, a more accurate expression for scattering function was used to include ionized state effects \cite{Hubbell}
\begin{equation}
	%	SF(K,N)=\bigg\langle\Psi\bigg|\sum^{N}_{i=1}\sum^{N}_{j=1}exp[i\mathbf{K}\cdot(\mathbf{r}_{i}-\mathbf{r}_{j})]\bigg|\Psi\bigg\rangle
	%	-[F(K,N)]^{2} \label{Waller Hartree_SF}
	SF(K,N)=\bigg\langle\Psi\bigg|\sum^{N}_{i=1}\sum^{N}_{j=1}e^{i\mathbf{K}\cdot(\mathbf{r}_{i}-\mathbf{r}_{j})}\bigg|\Psi\bigg\rangle-[F(K,N)]^{2} \label{Waller Hartree_SF}
\end{equation}
where the summation is over all electrons in atomic or molecular systems. In this equation, $|\Psi\rangle$ is the ground state for atomic or molecular systems, $\boldsymbol{r}_{i}$ and $\boldsymbol{r}_{j}$ are position vectors for $i$-th and $j$-th electrons correspondingly, and $F(K,N)$ is the total form factor for atoms or molecules defined by
\begin{equation}
	%	F(K,N)=\int\rho(r)exp(i\mathbf{K}\cdot\mathbf{r})d\mathbf{r} \label{Form Factor}
	F(K,N)=\int\rho(r)e^{i\mathbf{K}\cdot\mathbf{r}}d\mathbf{r} \label{Form Factor}
\end{equation}

The above results predicted by ISF approach are valid when variable $x$ and momentum transfer $K$ are large, which has been confirmed by experiments \cite{Kane1997,Pratt2004,Kane2006}. However, there are limitations on the ISF approach. Firstly, the ISF approach based on equation (\ref{SF differential}) can only be used to calculate the angular distribution for Compton scattering process. The information coming from the more differential quantities, such as DDCS $d^{2}\sigma/d\omega_{f}d\Omega_{f}$, are lost in the integration. Secondly, in the \emph{ab initio} calculations of scattering function $SF(\omega_{i},\theta)$, contributions from Compton scattering and Raman scattering cannot be efficiently distinguished \cite{Pratt,Chatterjee}. In numerical tabulations, contributions of these two processes are summed over to give a total result. In the high energy region, it is lucky the Compton scattering is dominant comparing with the Raman scattering.

The same as Compton profile $J(p_{z})$ discussed in section \ref{sec:3}, the scattering function $SF(\omega_{i},\theta)$ could also reflects the properties of target materials. Therefore, scattering function also opens up a bridge between Compton scattering and interdisciplinary studies. It offers opportunities to learn the electronic structure and properties of target materials from Compton scattering process \cite{Hubbell,Kane2006}.

\section{Scattering Matrix Approach \label{sec:5}}

In this section, we specialize in the scattering matrix (SM) approach. In this approach, atomic binding effects and many-body effects in atomic or molecular systems are taken into account more precisely. It not only provides us a more complete understanding of the main features of Compton scattering with bound electrons, but also helps us recognize the validity regions for other methods (such as FEA and IA). With these advantages, the SM approach has attracted large interests in interdisciplinary studies \cite{Pratt2,Drukarev2}.

In the SM approach, the DDCS of Compton scattering process can be calculated by the scattering matrix of a many-body theory
\begin{equation}
	\bigg( \frac{d^{2}\sigma}{d\omega_{f}d\Omega_{f}} \bigg)_{\text{SM}} \propto \mid M_{if} \mid^{2} \label{SM expression}
\end{equation}
Therefore, it is named as the scattering matrix (SM) approach. In this approach, the scattering matrix of Compton scattering process $M_{if} \propto \langle\Psi_{f}|H_{I}|\Psi_{i}\rangle$ can be calculated through the many-body interaction Hamiltonian $H_{I}$ for atomic or molecular systems. The interaction Hamiltonian $H_{I}$ is determined by electromagnetic interactions between bound electrons and radiation photon fields, and it is given by the Quantum Electrodynamics (QED) for atomic and molecular systems. There are two categories of SM approaches: the nonrelativistic and relativistic theories. In the nonrelativistic theories, the interaction Hamiltonian $H_{I}$ is expressed as
\begin{equation}
	H_{I}^{\text{nonrelativistic}} = \sum_{i=1}^{N}
	\bigg[
	\frac{e^{2}}{2m_{e}c}\boldsymbol{A}^{2} - \frac{e}{m_{e}c}\boldsymbol{p}_{i}\cdot\boldsymbol{A}
	\bigg] \label{Nonrelativistic Hamiltonian}
\end{equation}
Here, $\boldsymbol{A}$ is quantized electromagnetic vector potential which describe radiation field of incoming photon acting on atomic or molecular system. The $\boldsymbol{p}_{i}$ is the momentum for $i$-th electron, and the summation is over all electrons. In the relativistic theories, the interaction Hamiltonian is written by
\begin{equation}
	H_{I}^{\text{relativistic}} = -\sum_{i=1}^{N}c\boldsymbol{\alpha}_{i}\cdot\boldsymbol{A}
\end{equation}
where $\boldsymbol{\alpha}_{i}$ is the conventional Dirac matrices for $i$-th electrons.

The earlier works on SM approach were carried out based on the nonrelativistic Hamiltonian $H_{I}$, and they were restricted to $\boldsymbol{A}^{2}$ term (only the first term in the square bracket of equation (\ref{Nonrelativistic Hamiltonian}) was included) \cite{Eisenberger1,Schnaidt,Gummel,Schumacher}. The contributions from the second term $\boldsymbol{p}_{i}\cdot\boldsymbol{A}$ were accomplished by M. Gavrila \emph{et al.} in 1970s \cite{Gavrila1972a,Gavrila1972b,Gavrila1974,Gavrila1975}. The full relativistic treatment was first attempted by I. B. Whittingham \cite{Whittingham1,Whittingham2} and then developed by P. M. Bergstrom, T. Suri\'c, R. H. Pratt \emph{et al.} in 1990s \cite{Pratt,Pratt1,Bergstrom,Pratt1991,Suric1992}. In these works, the initial and final states ($|\Psi_{i}\rangle$ and $|\Psi_{f}\rangle$) in atomic or molecular systems were calculated within the independent particle model (IPM), in which the single-electron states are solved through the unperturbed Hamiltonian
\begin{equation}
	H_{0} = \frac{\boldsymbol{p}_{i}^{2}}{2m_{e}}+U(\boldsymbol{r}_{i}) \label{unperturbed Hamiltonian}
\end{equation}
Here, $\boldsymbol{p}_{i}$ and $\boldsymbol{r}_{i}$ are the momentum and position of the $i$-th electron. In the IPM method, the potential for $i$-th electron $U(\boldsymbol{r}_{i})$ is given by the mean field of atomic nuclear potential and many-body interactions from other bound electrons. The IPM treatment may not be a perfect method in dueling with electron non-local exchange and correlations, and new treatments to tame electron non-local exchange in the framework of SM is still in development \cite{Chatterjee}. With the initial and final state wavefunctions $|\Psi_{i}\rangle$, $|\Psi_{f}\rangle$, and interaction Hamiltonian $H_{I}$, the DDCS of Compton scattering process can be achieved from equation (\ref{SM expression}). There are a lot of techniques to calculate the matrix elements $M_{if}$ in equation (\ref{SM expression}), which is beyond the scope of this work. In particular, in the nonrelativistic theories, the scattering matrix element $M_{if} \propto \langle\Psi_{f}|H_{I}|\Psi_{i}\rangle$ reduces to the Kramers-Heisenberg-Waller (KHW) matrix element \cite{Kramer,Waller1928a,Waller1928b}.

\begin{figure}
	\includegraphics[width=0.525\textwidth]{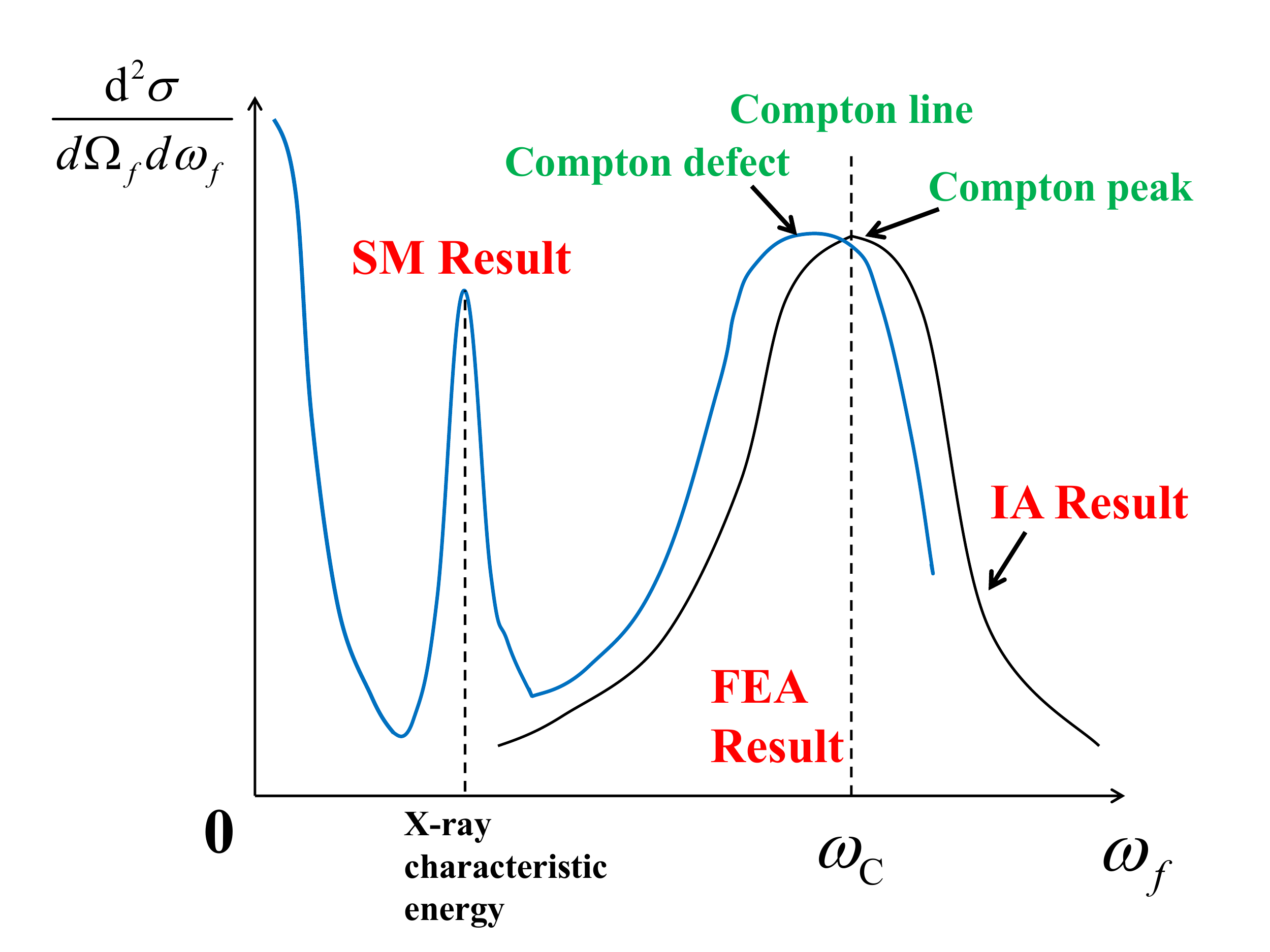}
	\caption{Schematic plot of Compton scattering DDCS in the SM approach at the scattering angle $\theta$. The horizontal axis labels the finial photon energy $\omega_{f}$, and the vertical axis labels the DDCS of Compton scattering process.}
	\label{Compton Scattering Distribution SM}
\end{figure}

The spectrum of DDCS in Compton scattering obtained from SM approach has three main features. Firstly, when final state photon energy approach to zero, namely in the $\omega_{f} \rightarrow 0$ limit, an infrared rise emerges in the spectrum \cite{Pratt2}. This infrared rise is an example of infrared divergence behavior, which is a common feature of QED \cite{Sakurai,Peskin}. The same infrared rise behavior is also predicted in the low-energy theorem (LET) method \cite{Rosenberg1991,Gavrila2000} 
\footnote{Infrared behavior arises in the $\omega_{f} \rightarrow 0$ limit, and the energy transfer $T=\omega_{i}-\omega_{f} \rightarrow \omega_{i}$. In these cases, the energy of outgoing photons is tiny, and it makes these outgoing photons very difficult to observe experimentally. The Compton scattering process is happened as if the incident photon energy $\omega_{i}$ is totally absorbed by atomic or molecular systems without producing notable influences, similar to the photoionization process / photoelectric effect. }. 
Secondly, some resonant peaks appear near the characteristic X-rays energies, which are the transition energies between different atomic or molecular states 
\footnote{When the resonant peaks emerge at X-ray characteristic energies, the electron undergoes a deexcitation process to lower energy intermediate states before it is ionized in Compton scattering process. The resonant peak energies correspond to the transition energies in these deexcitation subprocesses.}.  
This resonant behavior mostly come from Compton scattering with L and M shell electrons \cite{Pratt}. At last, the ``Compton peak'' is reproduced in the vicinity of Compton energy $\omega_{C}$. In the SM results, the center of Compton peak $\omega_{\text{cen}}^{\text{SM}}$ is slightly different from the Compton energy $\omega_{C}$, and the difference $\delta=\omega_{\text{cen}}^{\text{SM}}-\omega_{C}$ leads to the ``Compton defect'' or ``asymmetry of Compton profile'' \cite{Pratt2,Chatterjee,Chatterjee2007} 
\footnote{There are other ways to define ``asymmetry of Compton profile''. For instance, in reference \cite{Chatterjee2007}, the ``asymmetry of Compton profile'' is defined as $A=\frac{J(p_{z},K)-J(-p_{z},K)}{J(p_{z}=0,K)}$. Here, $K$ is the modulus of momentum transfer vector $\boldsymbol{K} \equiv \boldsymbol{k}_{f}-\boldsymbol{k}_{i}$ in Compton scattering process, and $J(p_{z},K)=\frac{(d^{2}\sigma/d\Omega_{f}d\omega_{f})_{\text{SM}}}{Y^{\text{IA}}}$ is the effective Compton profile extracted from SM results. Actually, from a detailed analysis of kinematical behavior in Compton scattering process, the following conclusions can be obtained. If the summit of Compton peak in the SM result is located at Compton energy $\omega_{C}$ exactly, namely $\delta=\omega_{\text{cen}}^{\text{SM}}-\omega_{C}=0$, the extracted Compton profile $J(p_{z},K)$ would reach its maximum value at $p_{z}=0$. In this case, $J(p_{z},K)$ is axi-symmetric around $p_{z}=0$ for spherical symmetric atoms, the same as conventional Compton profile in equation (\ref{electron profile2}), and there is no ``asymmetry''. However, when the summit of Compton peak is shifted from Compton energy $\omega_{C}$, $J(p_{z},K)$ would not get the maximal value at $p_{z}=0$. In this case, $J(p_{z},K)$ is not axi-symmetric around $p_{z}=0$ for spherical symmetric atoms, so there is asymmetry in $J(p_{z},K)$ between positive and negative $p_{z}$ values.}. 
When the modulus of momentum transfer $K$ in Compton scattering is sufficiently large such that inequalities $p_{\text{averge}}/K \ll 1$ and $a/K \ll 1$ are satisfied, the ``Compton defect'' becomes extremely small and it can be neglected \cite{Pratt1,Pratt2,Pratt,Bergstrom,Suric1992,Chatterjee2007}. Here, $p_{\text{averge}}$ is the average momentum in atomic or molecular systems, $a$ is parameter defined to be $a=m_{e}cZ\alpha$, and $\alpha \approx 1/137$ is the fine-structure constant. In these cases, the RIA result does not present notable deviations from the SM result in the Compton peak region. The spectrum of DDCS predicted by SM result is illustrated in figure \ref{Compton Scattering Distribution SM}. To summarize, in the SM results, DDCS of Compton scattering exhibit infrared rise, the resonant peak, and the Compton peak. The three categories of peaks arise in different energy ranges. When we discuss the Compton peak, the final photon energy $\omega_{f}$ should near Compton energy $\omega_{C}$, and the incident photon energy $\omega_{i}>\omega_{C}$. When the resonant peak is observed, final photon energy $\omega_{f}$ is just near the X-ray characteristic energy of atoms or molecules, and initial photon energy $\omega_{i}$ is usually much larger than this characteristic energy. In order to see the infrared rise, final photon energy $\omega_{f}$ should goes to zero (namely $\omega_{f} \rightarrow 0$), while the incident energy $\omega_{i}$ is not necessary to be very small.

It should be noted that the SM approach is still in development now \cite{Pratt2,Drukarev1,Drukarev2,Chatterjee,Jung1998,Pratt2000,Hopersky1,Hopersky2,Hopersky3}. In some studies, new treatments and techniques are pursued to handle electron non-local exchange and correlation by methods beyond IPM \cite{Jung1998,Pratt2000,Hopersky1}. Other studies are devoted to more complex scattering process, for instance, A. N. Hopersky \emph{et al.} investigated the Compton scattering and Rayleigh scattering of two X-ray photons \cite{Hopersky2,Hopersky3}. Apart from the theoretical explorations, there are several experiments which have provided evidences to confirm the SM approach \cite{Sparks,Kane1997,Jung1998,Pratt2000,Pratt2004,Chatterjee,Kircher}. Recently, Max Kircher \emph{et al.} conducted a kinematically complete Compton scattering experiment utilizing X-rays produced from accelerators with energy about 2.1 keV. By measuring the angular distribution of the scattered photon, the experimental observations present large deviations with the FEA results, but the experimental data are consistent with theoretical predictions from SM approach \cite{Kircher}. This observation indicates that the SM approach is becoming a promising tool to duel with Compton scattering with bound electrons. Furthermore, the resonant peaks in Compton spectrum near the characteristic X-rays energies predicted by SM approach have also been confirmed by experiments \cite{Sparks}. However, despite lots of attempts, the infrared rise behavior in Compton spectrum predicted by SM approach (which is also predicted by LET approach) has not been confirmed in experiments yet.

Although there is still inadequacy in duel with some many-body effects (such as the electron non-local exchange), the SM is the most advanced and accurate approach in the \emph{ab initio} calculations of Compton scattering with bound electrons over the past years. Firstly, in SM approach, the initial state $|\Psi_{i}\rangle$, final state $|\Psi_{f}\rangle$, and the dynamical process of Compton scattering are all treated by many-body QED theory of atomic or molecular systems. It is a fully quantum many-body approach, not just making simple corrections to the FEA results, like RIA and ISF approaches. Secondly, the SM approach can take the many-body effects in atomic or molecular systems into account as much as possible. In this approach, atomic bindings, electron motions around atomic nuclei and electron many-body interactions are all considered in the starting points of theoretical treatments. Thirdly, SM results can reflect all the main features in Compton scattering process: the infrared rise at low energy, the resonant peak at X-ray characteristic energy, and the broaden ``Compton peak'' near the Compton energy $\omega_{C}$. With the aforementioned superiority, SM can make more accurate predictions in the entire region of the spectrum. Furthermore, many other approaches, such as FEA and IA, can be derived from the SM approach by making appropriate and simplified approximations. Since it was developed, SM have attracted lots of interests in atomic and molecular physics, and it may has great impacts in these areas in the near future.

\section{Comparisons between Theoretical Calculations and Experimental Measurements \label{sec:6}}

In the section, in order to have a better understanding of characteristics and limitations of the approaches described above, the comparisons between theoretical calculations and experimental measurements are provided. Limited by the scope of this work, only a few representative examples are presented. For more examples, the readers can resort to references \cite{Pratt1,Kane1992,Kane1997,Kane2006,Kurucu,Yalcin,Wang2020,Rullhusen1976,Basavaraju1987,Manninen1990}.

\begin{figure}
\includegraphics[width=0.525\textwidth]{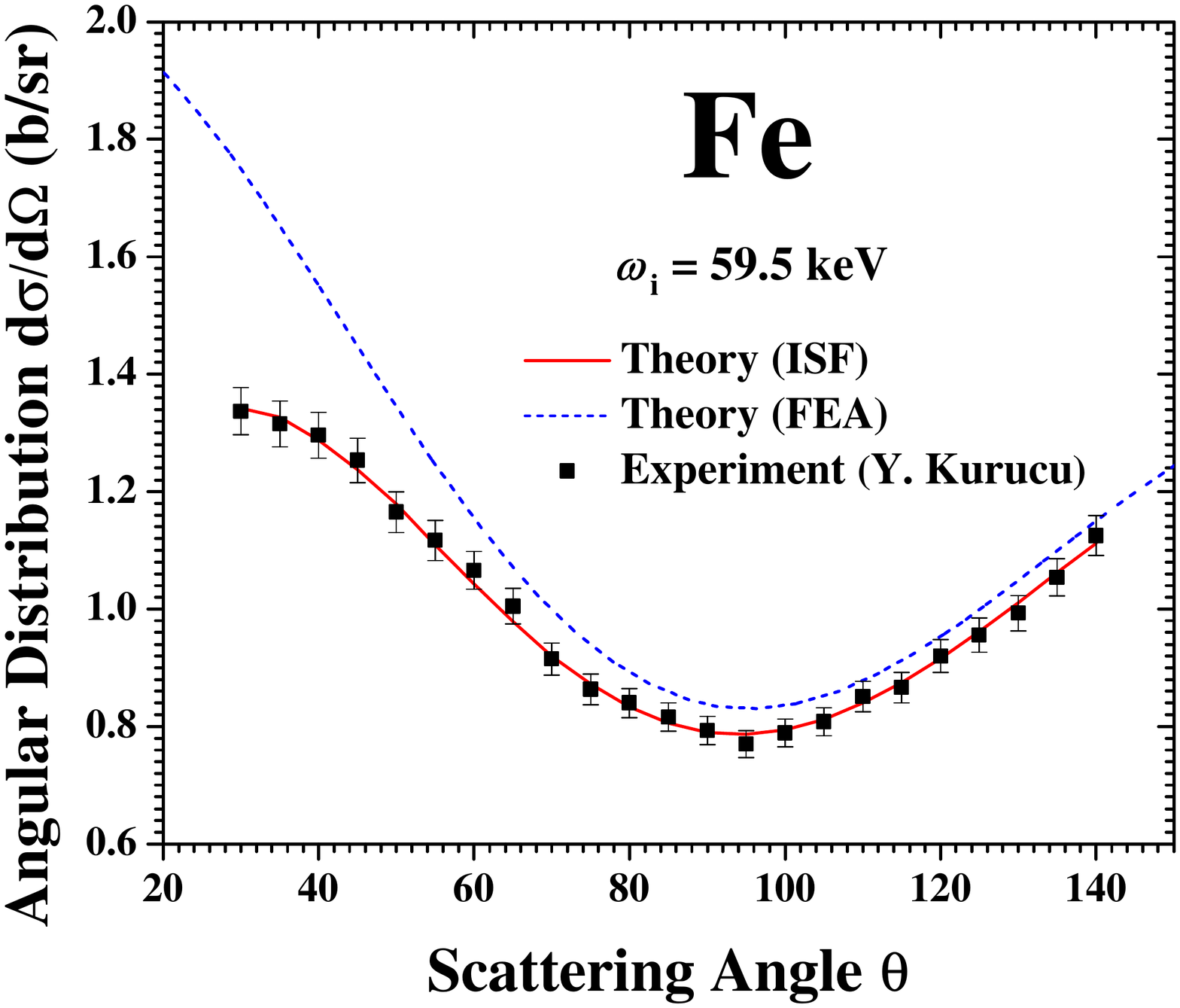}
\includegraphics[width=0.525\textwidth]{Figure4b_new3.pdf}
\caption{Angular distribution for Compton scattering process at incident photon energy $\omega_{i}=59.5$ keV for Fe and Cu. In this figure, the vertical axis labels differential cross sections $d\sigma/d\Omega_{f}$, and the horizontal axis labels the scattering angle $\theta$. The experimental measurements from reference \cite{Kurucu,Yalcin} and the theoretical predictions of FEA and ISF approaches \cite{Yalcin,Wang2020} are plotted in this figure.}
\label{Theory and Experiment}
\end{figure}

For the angular distribution of Compton scattering process, the theoretical and experimental results of Fe and Cu are presented in figure \ref{Theory and Experiment}. In this figure, the incident photon energy is $\omega_{i}=59.5$ keV. We choose Fe and Cu elements as representative examples of  elemental metal and ferromagnetic metal, respectively. Results of other elements exhibit a similar behavior, and they are not displayed in this figure. The readers could find more examples in references \cite{Kurucu,Yalcin,Wang2020,Kane1992}. The experimental measurements from references \cite{Kurucu,Yalcin} and the theoretical predictions of FEA and ISF approaches\cite{Yalcin,Wang2020} are plotted in this figure. In reference \cite{Wang2020}, the scattering function $SF(\omega_{i},\theta)$ is calculated using equation (\ref{SF RIA}) in RIA framework \cite{Yalcin,Wang2020} 
\footnote{Note that in section \ref{sec:4}, we have mentioned that the scattering function $SF(\omega_{i},\theta)$ can be obtained from RIA, Waller-Hartree theory, and other methods. In particular, within the RIA framework, the scattering function $SF(\omega_{i},\theta)$ is calculated through the integration of Compton profile $J(p_{z})$, see equations (\ref{SF RIA})-(\ref{electron profile}).}. 
The results in figure \ref{Theory and Experiment} indicate that ISF is a better approach than FEA in calculating the angular distribution for Compton scattering process. The ISF results successfully reproduce experimental measurements at all angles, while the FEA results can bring about large discrepancies in the small angle regions. In this region, both the energy transfer and momentum transfer in Compton scattering process are very small, so that atomic electrons cannot be viewed as free electrons anymore. In the low-energy and low-momentum transfer region (correspond to small scattering angle $\theta$), only a small percentage of bound electrons in atomic or molecular systems are activated in the Compton scattering process. Therefore, the FEA results, in which all electrons are treated as free and activated, would tremendously overestimate the angular distribution of Compton scattering process. 

\begin{figure*}
\includegraphics[width=0.57\textwidth]{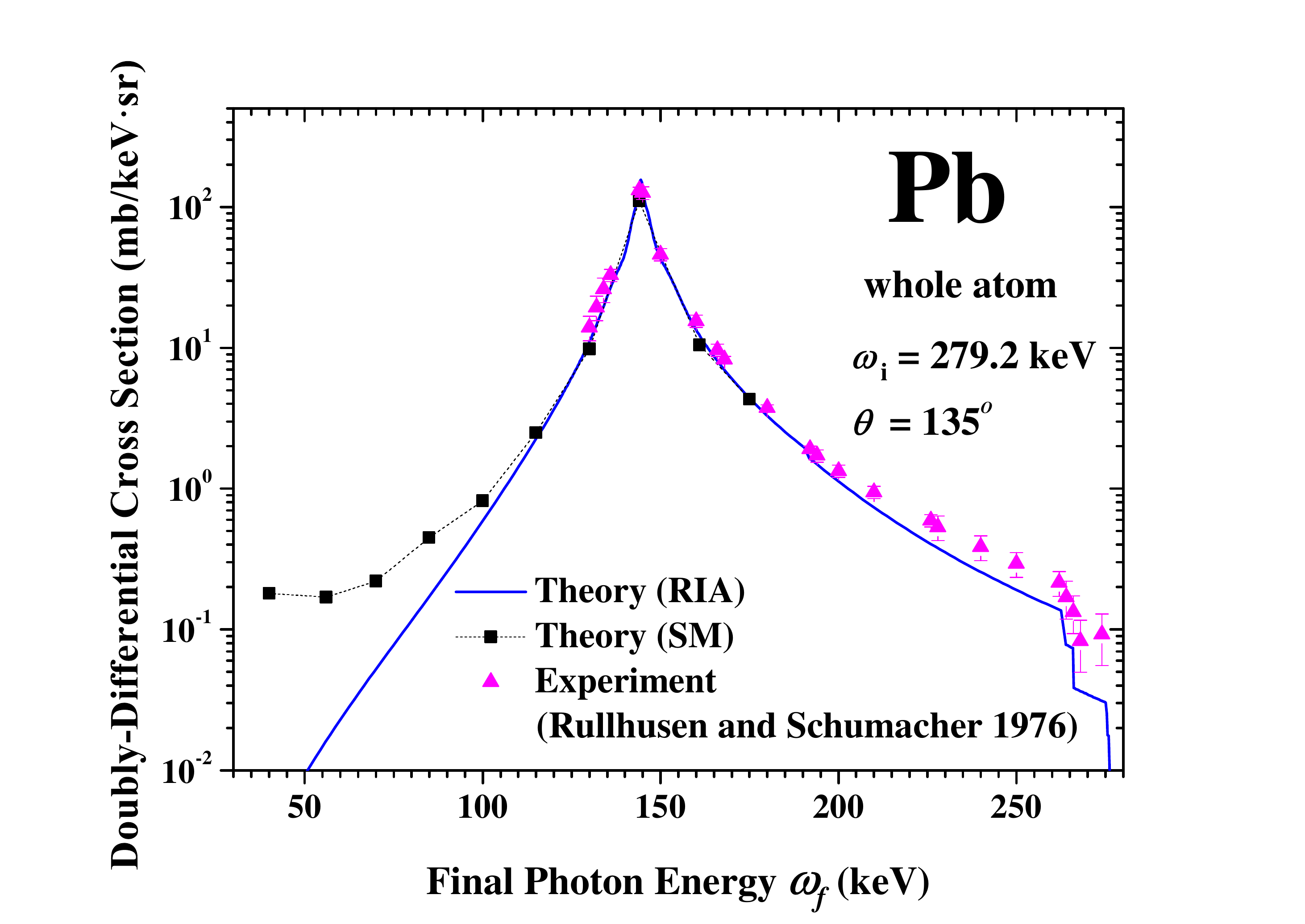}
\includegraphics[width=0.57\textwidth]{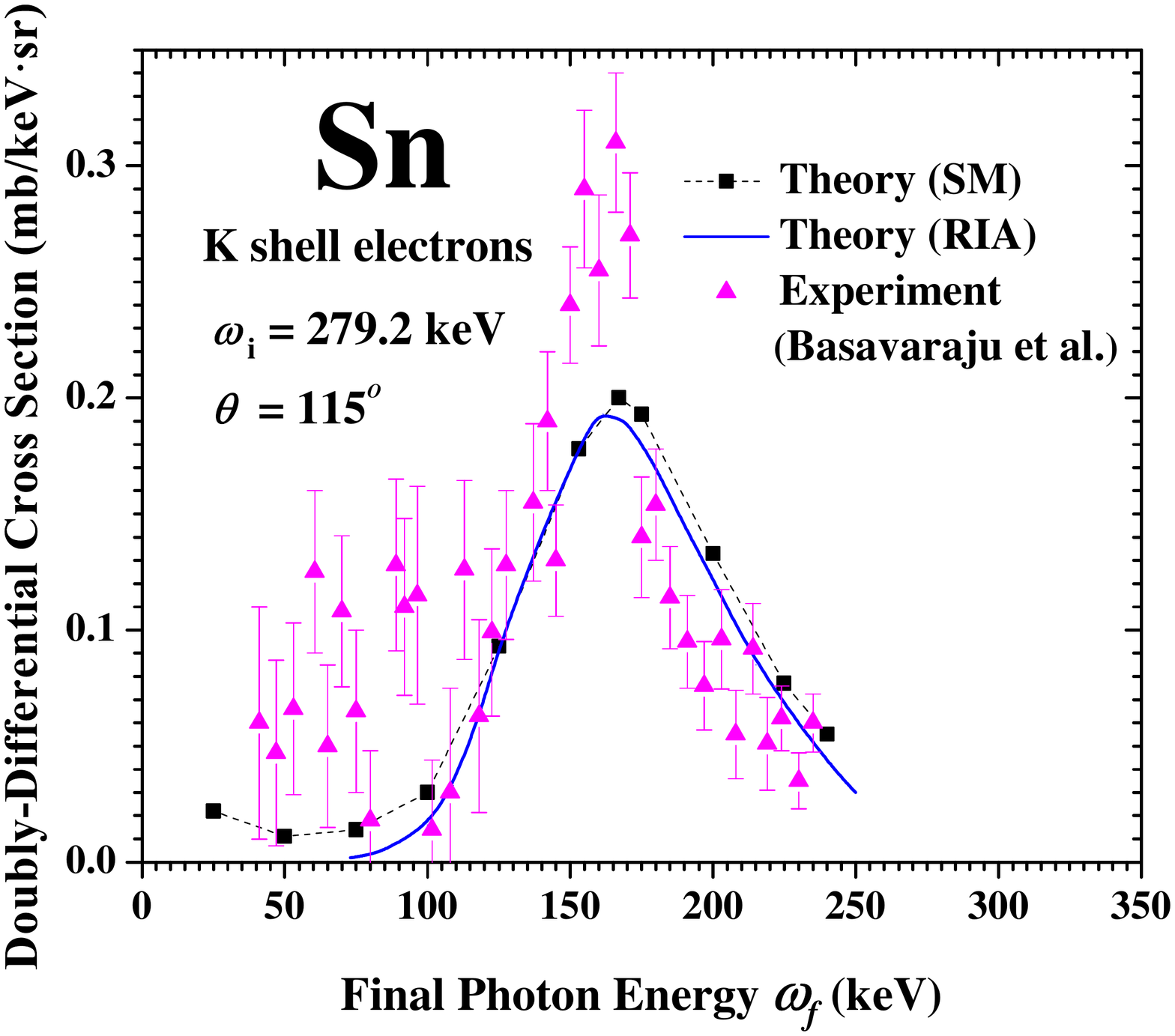}
\includegraphics[width=0.57\textwidth]{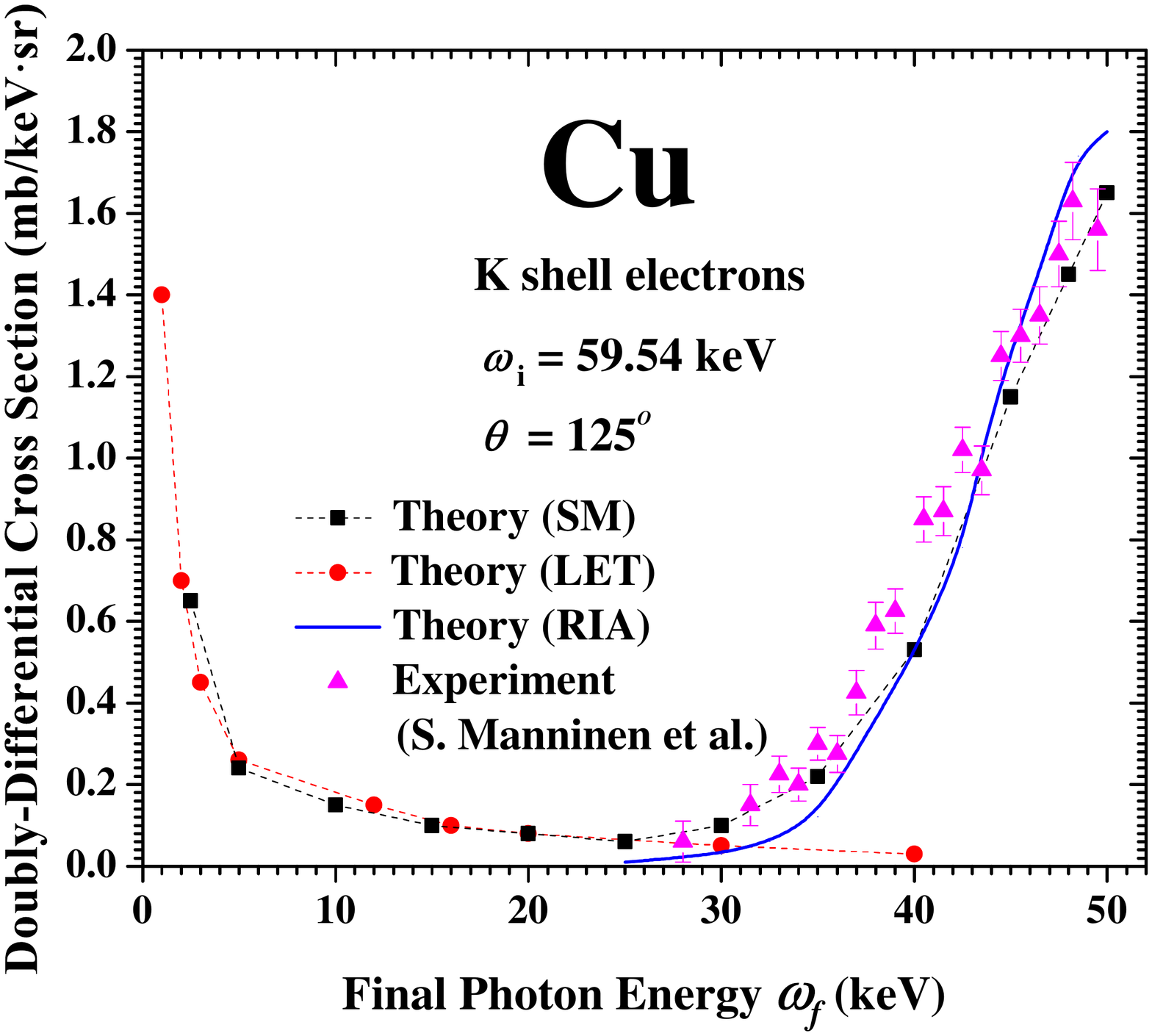}
\caption{DDCS of Compton scattering for Cu and Pb elements. In this figure, the vertical axis labels the DDCS $d\sigma/d\Omega_{f}d\omega_{f}$, and the horizontal axis labels the final photon energy $\omega_{f}$. The subfigures (a), (b) and (c) correspond to the following cases: (a). Compton scattering for Pb atom at $\omega_{i}=279.2$ keV and $\theta=135^{o}$; (b). Compton scattering for K-shell electrons of Sn atom at $\omega_{i}=279.2$ keV and $\theta=115^{o}$; (c). Compton scattering for K-shell electrons of Cu atom at $\omega_{i}=59.5$ keV and $\theta=125^{o}$. The experimental measurements are given by references \cite{Rullhusen1976,Basavaraju1987,Manninen1990}. The theoretical predictions from RIA and SM approaches are given by Qiao \emph{et al.} and Bergstrom \emph{et al.} \cite{Pratt1,Qiao,Bergstrom}.}
\label{Theory and Experiment2}
\end{figure*}

For the DDCS of Compton scattering process, the theoretical and experimental results are presented in figure \ref{Theory and Experiment2}. This figure gives the results of Cu, Sn and Pb elements. It is worth noting that, for the DDCS, the FEA result becomes singular at the Compton energy $\omega_{C}$, while it gets zero at other energies. This is due to the Dirac delta function $\delta(\omega_{f}-\omega_{C})$ in equation (\ref{KN doubly-differential}). For this reason, it is not that valuable to plot the FEA result, what we need to focus are the IA and SM predictions. In this figure, subfigures (a), (b) and (c) correspond to the following cases: (a). Compton scattering for Pb atom at $\omega_{i}=279.2$ keV and $\theta=135^{o}$; (b). Compton scattering for K-shell electrons of Sn atom at $\omega_{i}=279.2$ keV and $\theta=115^{o}$; (c). Compton scattering for K-shell electrons of Cu atom at $\omega_{i}=59.5$ keV and $\theta=125^{o}$. The subfigures (a) and (b) display the Compton peak region, while the subfigure (c) presents the spectrum outside the Compton peak. In the subfigure (a), the SM result is calculated by Bergstrom \emph{et al.} \cite{Bergstrom}, the RIA result is given by Qiao \emph{et al.} \cite{Qiao}, and experimental measurements are given by Rullhusen and Schumacher \cite{Rullhusen1976}. In the subfigures (b) and (c), the theoretical SM results are given through combined works of Bergstrom \emph{et al.} and Gavrila \cite{Pratt1,Gavrila1974}, the RIA result is given by Bergstrom \emph{et al.} \cite{Pratt1}, and the experimental measurements are given by Basavaraju \emph{et al.} \cite{Basavaraju1987} and Manninen \emph{et al.} \cite{Manninen1990}, respectively. In subfigure (c), the LET results are calculated by Bergstrom \emph{et al.} \cite{Pratt1}. From these comparisons, some conclusions can be drawn. Firstly, for the DDCS of Compton scattering, the FEA approach becomes deficient and inconvenient, because of the singular behavior in the spectrum. Secondly, the SM and RIA results are similar in Compton peak region $\omega_{f}\approx\omega_{C}$. Both RIA and SM results are consistent with experimental observations in the Compton peak region, when some discrepancies are included. Thirdly, the SM result is largely different from the RIA result outside the Compton peak, due to the infrared rise mentioned in section \ref{sec:5} (and possible resonant peaks near X-ray characteristic energies, which are not emerged in subfigure (c) \footnote{In the SM results, resonant peak behavior often appears in the cases of L and M shell electrons \cite{Pratt,Pratt1}. In figure \ref{Theory and Experiment2}, subfigure (c) corresponds to the Compton scattering with K shell electrons, so the resonant peaks do not emerge.}). In regions far from the Compton peak, especially the infrared region where final photon energy $\omega_{f}$ is very small, more experimental data are required to test the SM results.

Since RIA result is reliable and consistent with experimental observations in the Compton peak region $\omega_{f}\approx\omega_{C}$, we can safely use Compton profile $J(p_{z})$ in IA approach to tackle Compton scattering in peak region. In the past decades, many researches emerged to study Compton profiles using the theoretical IA approach combined with experimental measurements near Compton peak \cite{Cooper3,Rathor,Pisani}. In experiments, the Compton profiles for atomic or molecular systems can be exacted from experimental data via equation
\begin{equation}
	\bigg[J(p_{z})=\iint\rho(\boldsymbol{p})dp_{x}dp_{y}\bigg]_{\text{exp}}=\frac{\big(\frac{d^{2}\sigma}{d\Omega_{f}d\omega_{f}}\big)_{\text{exp}}}{Y^{\text{IA}}}
\end{equation}
Here, $\big(d^{2}\sigma/d\Omega_{f}d\omega_{f}\big)_{\text{exp}}$ is the experimental measured DDCS of Compton scattering. Experimental studies for Compton profile has attracted huge interests in recent years \cite{Cooper3,Aguiar2015,Suortti1999}. Furthermore, there are other quantities similar to the conventional Compton profile $J(p_{z})=\iint\rho(\boldsymbol{p})dp_{x}dp_{y}$ discussed above. For example, if the incident photon beams are polarized, the differential cross-section is connected with the magnetic Compton profile $[J(p_{z})]_{\text{mag}}=\iint[\rho^{\uparrow}(\boldsymbol{p})-\rho^{\downarrow}(\boldsymbol{p})]dp_{x}dp_{y}$ \cite{Sakurai2004,Benea2018,Agui2019,Dashora2020,James2021}, where $\rho^{\uparrow}(\boldsymbol{p})$ and $\rho^{\downarrow}(\boldsymbol{p})$ are the spin polarized electron momentum densities. Other kinds of Compton profiles, such as the directional Compton profile obtained by setting $z$ axis along different crystallographic axes, are also widely studied in recent years \cite{Cooper3,Apell2000,Koizumi2019}. Limited to the scope of the present work, we only focus ourselves on the conventional Compton profile $J(p_{z})$ defined in equation (\ref{Compton profile}). Other kinds of Compton profiles are not discussed in details. More information on various kinds of Compton profiles can be found in references \cite{Cooper3,Sakurai2004}.

\section{Database and Applications \label{sec:7}}

In the past several decades, Compton scattering had been extensively applied into many branches of science, including atomic \cite{Pratt}, molecular, condensed matter \cite{Cooper3,Kubo,Rathor,Wang,Pisani}, astrophysical \cite{Porter}, nuclear and elementary particle physics \cite{Brusa,Ji,Ramanathan}. A lot of experimental and theoretical investigations concerning X-rays and gamma-rays cannot be carried on without the help of Compton scattering \cite{Kane1997,Kane1992,Monash}. As discussed in section \ref{sec:3} and section \ref{sec:4}, Compton scattering is a powerful tool to study momentum distribution for bound electron in atomic, molecular and condensed matter systems \cite{Cooper1,Cooper2,Cooper3}, both for theoretical and experimental studies. With the help of Compton profile and Compton scattering experiments, electron correlations, Fermi surfaces and band structures in materials can be investigated \cite{Kubo,Pisani,Rathor,Wang}. The Compton profiles are also closely related to the positron annihilation angular correlation spectra \cite{Dugdale2014}. Furthermore, the development of modern gamma-ray spectrometer and imaging devices is also benefits a lot from the Compton scattering \cite{Takada,Mihailescu,Chiu,Phuoc}.

In the passed years, many databases on Compton scattering had already been built up. The most common quantities in tabulations and databases are Compton profile $J(p_{z})$, incoherent scattering function $SF(\omega_{i},\theta)$, and total cross section $\sigma$ 
\footnote{In the theoretical calculations on scattering function $SF(\omega_{i},\theta)$ and total cross section $\sigma$, contributions from Compton scattering and Raman scattering cannot be fully distinguished \cite{Pratt}. In numerical tabulations, the contributions from these two processes are summed over to give a total result. In the high energy region, the Compton scattering process is dominant compared with the Raman scattering.}. 
For engineering or industrial applications, the data of differential cross sections $d^{2}\sigma/d\omega_{f}d\Omega_{f}$ and $d\sigma/d\Omega_{f}$ in Compton scattering can be easily achieved from the tabulation of Compton profile and scattering function (using equations (\ref{IA}) and (\ref{RIA2})). The most widely used database on Compton profile $J(p_{z})$ over past years was given by F. Biggs \emph{et al.} in 1975, in which a complete study on atomic Compton profile for elements ($1\leq Z\leq 102$) was presented \cite{Biggs}. In Biggs's work, the nonrelativistic Hartree-Fock theory was used to calculate Compton profiles for light elements ($1\leq Z\leq 36$) and the relativistic Dirac-Hartree-Fock theory was used to calculate Compton profiles for heavy elements ($36\leq Z\leq 102$). On the scattering function $SF(\omega_{i},\theta)$, J. H. Hubbell \emph{et al.} provided extensive and widely available tabulations for elements ($1\leq Z\leq 100$) based on equation (\ref{Waller Hartree_SF}) in 1975, with ground state wavefunctions calculated by several methods \cite{Hubbell}. Later, S. Kahane gave refined calculations using RIA approach and Dirac-Hartree-Fock ground state wavefunctions for all elements ($1\leq Z\leq 110$) in 1998 \cite{Kahane}. For the total cross section $\sigma$, J. H. Hubbell \emph{et al.} also gave tabulations for elements ($1\leq Z\leq 100$) in their early work in 1975 \cite{Hubbell}. The up-to-date tabulations on total cross section $\sigma$ are provided by EPDL and NIST databases, which combine the theoretical and experimental data \cite{EPDL97,NIST XCOM}.

\section{Summary \label{sec:8}}

Throughout this paper, we give an overview of the theoretical approaches on the \emph{ab initio} calculation for Compton scattering with bound electrons in atomic or molecular systems. In this work, we focus on the basic ideas and main results for several approaches. The advantages, validity ranges, and applications of each approach are also briefly explained. These approaches are free electron approximation (FEA), impulse approximation (IA), incoherent scattering factor / incoherent scattering function approximation (ISF), scattering matrix (SM). Limited to the scope of this work, other approaches and applications are not discussed here, and there are many important works we have not mentioned in this work. We hope that this work would be helpful to theoreticians and experimentalists, especially for those who work on interdisciplinary branches of science with the help of Compton scattering.
	
\section*{ACKNOWLEDGMENTS}

The authors should also thank to the great efforts from all around the world during the pandemic period of Covid-19. This work was supported by the Scientific Research Foundation of Chongqing University of Technology (Grants No. 2020ZDZ027 and No. 2019ZD21), the Natural Science Foundation of Chongqing (Grant No. 2020CCZ036).

\section*{ABBREVIATIONS}

The following abbreviations are used in this work:\\
\begin{ruledtabular}
\noindent 
\begin{tabular}{@{}ll}
	DDCS & Doubly-differential Cross Section\\ 
	DHF & Dirac-Hartree Fork\\ 
	FEA & Free Electron Approximation\\ 
	HF & Hartree Fork\\ 
	IA & Impulse Approximation\\ 
	NRIA & Nonrelativistic Impulse Approximation\\ 
	RIA & Relativistic Impulse Approximation\\ 
	ISF & Incoherent Scattering Function \\
	& Incoherent Scattering Factor\\ 
	SM & Scattering Matrix\\
	LET & Low Energy Theorem
\end{tabular}
\end{ruledtabular}

\

\end{document}